 \documentstyle[prl,aps]{revtex}  

\newcommand{\tsub}[1]{_{\mbox{\scriptsize#1}}}

\newcommand{\tfrac}[2]{\mbox{$\small\frac{#1}{#2}$}}
\newcommand{\quarterthin}{\kern 0.0417em}

\newcommand{\bra}[1]{\langle#1|}
\newcommand{\ket}[1]{|#1\rangle}
\newcommand{\ev}[1]{\langle#1\rangle}
\newcommand{\mel}[3]{\bra{#1}#2\ket{#3}}
\newcommand{\thin}{\thinspace}

\begin{document}
\draft


 \twocolumn[\hsize\textwidth\columnwidth\hsize  
 \csname @twocolumnfalse\endcsname              

\title
{An SU(4) Approach to High-Temperature Superconductivity and
Antiferromagnetism }

\author
{Mike Guidry$^{(1)}$, Lian-Ao Wu$^{(2)}$, Yang Sun$^{(1)}$ and Cheng-Li
Wu$^{(3)}$}
\address{
$^{(1)}$Department of Physics and Astronomy, University of Tennessee,
Knoxville, Tennessee 37996 \\
$^{(2)}$Department of Physics, Jilin University, Changchun, Jilin, 130023 PRC\\
$^{(2)}$Department of Physics, Chung-Yuan Christian University, Taiwan, ROC}

\date{\today}
\maketitle

\begin{abstract} 
We present an $SU(4)$ model of high-$T_c$ superconductivity.
One dynamical symmetry of this model corresponds to the 
previously proposed $SO(5)$ model for unification of superconductivity and
antiferromagnetism, but there are two additional dynamical symmetries: $SO(4)$,
associated with antiferromagnetic order and
$SU(2)$, associated with a D-wave pairing condensate.  These provide a 3-phase
microscopic model of high-$T_c$ superconductivity and permit a clear
understanding of the role played by the $SO(5)$ symmetry.
\end{abstract}

\pacs{}

 ]  

\narrowtext

There are strong arguments that the mechanism for high-temperature
superconductivity (SC) is not BCS S-wave pairing \cite{sca95}. The interaction
leading to the formation of the singlet pairs appears not to be the traditional
lattice phonon mechanism underlying the BCS theory but a collective electronic
interaction. The pairing gap has nodes in the $k_x$--$k_y$ plane suggestive of
D-wave hybridization in the 2-particle wavefunctions, and SC in
the cuprates seems to be closely related to the antiferromagnetic (AF) insulator
properties of their normal states.
Furthermore, the formation of Cooper pairs and their condensation 
in high-$T\tsub c$ compounds may be distinct, with pair formation at a
higher temperature scale than the formation of the SC state.  
This is reminiscent of grand unified theories in particle physics,
where a hierarchy of symmetry breakings
is implied by a Lie group structure that is broken spontaneously 
(and explicitly) on
different temperature scales. Such observations argue for a theory  
that is based on 
symmetries describing pairing more sophisticated than 
BCS and that can unify
phases such as AF order and SC in the same 
theory.

Dynamical symmetries
having many of these characteristics have been
developed extensively in nuclear
structure physics \cite{wu94,IBM}.
There it has proven fruitful to ask the following
questions: what are the relevant low-lying collective degrees of freedom,
what are the microscopic
operators
that create and annihilate these modes, and what is the commutator algebra
that they obey?
It has been shown that dynamical symmetries having the representation structure
described in Ref.\ \cite{wu94} are realized to remarkably high accuracy in the
spectrum and the wavefunctions of large-scale numerical calculations
\cite{sun98}.

Recently, S. C. Zhang
and others have applied related ideas
to high-temperature SC 
\cite{zha97,hen98,rab98,MV98}.
In order to unify AF and SC order
parameters, Zhang \cite{zha97} assembled these into a 5-dimensional vector order
parameter and constructed an $SO(5)$ group that rotates
AF order into SC order.
In this paper we proceed differently by closing a minimal fermion algebra
containing D-wave pairs of singlet spin 
(D-pairs) and triplet spin ($\pi$-pairs), charge, and spin operators. 
Nevertheless, we shall recover
an $SO(5)$ subgroup of a more general $U(4)$ symmetry.

Thus,
the recent discussion of Zhang's $SO(5)$ symmetry applies
directly to our results,
but we extend this
discussion in three ways:  (1)~The $SO(5)$ subgroup is embedded in a
larger fermion algebra that
constrains the $SO(5)$ subgroup.
(2)~The $SU(4)$ highest symmetry
has subgroups  in addition to $SO(5)$ that may be relevant for AF and SC 
phases, and that aid in
interpreting the $SO(5)$ symmetry.
(3)~We implement
an exact {\em dynamical
symmetry} using  Casimir invariants of group chains and a
corresponding collective subspace of low dimensionality.

Let us introduce bilinear fermion generators: 
\begin{eqnarray} 
p_{12}^\dagger&=&\sum_k g(k) c_{k\uparrow}^\dagger
c_{-k\downarrow}^\dagger
\qquad p_{12}=\sum_k g^*(k) c_{-k\downarrow} c_{k\uparrow} \nonumber
\\ q_{ij}^\dagger &=& \sum_k g(k) c_{k+Q,i}^\dagger c_{-k,j}^\dagger
\qquad q_{ij} = (q_{ij}^\dagger)^\dagger
\\ Q_{ij} &=& \sum_k c_{k+Q,i}^\dagger c_{k,j} \qquad S_{ij} = \sum_k
c_{k,i}^\dagger c_{k,j} - \tfrac12 \Omega \delta_{ij}  \nonumber
\end{eqnarray} 
where $c_{k,i}^\dagger$ creates a fermion of momentum $k$ and
spin projection $i,j= 1 {\rm\ or\ }2 = \ \uparrow$ or
$\downarrow$, $Q=(\pi,\pi,\pi)$ is an AF ordering vector,
$\Omega$ is the lattice degeneracy, and
$ g(k) = {\rm sgn} (\cos k_x -\cos k_y)
$ with $g(k+Q) = -g(k)$ and
$\left| g(k) \right| = 1$ (see Refs.\ \cite{hen98,rab98}).  
The operator
set (1) is closed under commutation, generating a  $U(4)$ Lie algebra 
with 3 dynamical symmetry chains:
\begin{eqnarray}
&\supset& SO(4) \times U(1) \supset SU(2)\tsub{s} \times U(1) \nonumber \\ U(4)
\supset SU(4) &\supset& SO(5) \supset SU(2)\tsub{s} \times U(1) 
\label{eq3}
\\ &\supset& SU(2)\tsub{p}
\times SU(2)\tsub{s} \supset SU(2)\tsub{s} \times U(1) \nonumber
\end{eqnarray} 
that end in the 
subgroup $SU(2)\tsub{s} \times U(1)$ representing spin and charge
conservation.  The physical interpretation is aided by 
rewriting the generators of the $U(4)$ algebra as
\begin{eqnarray} 
Q_+&=&Q_{11}+Q_{22} = \sum_k (c_{k+Q\uparrow}^\dagger
c_{k\uparrow} + c_{k+Q\downarrow}^\dagger c_{k\downarrow}) \nonumber
\\
\vec S &=& \left( \frac{S_{12}+S_{21}}{2},
                \ -i \, \frac {S_{12}-S_{21}}{2},
                \ \frac {S_{11}-S_{22}}{2} \right) \nonumber
\\
\vec {\cal Q} &=& \left(\frac{Q_{12}+Q_{12}}{2},\ -i\, \frac{Q_{12}-Q_{21}}{2},
\ \frac{Q_{11}-Q_{22}}{2} \right)
\\
\vec \pi^\dagger &=& \left(
\tfrac i2\, (q_{11}^\dagger - q_{22}^\dagger), \
\tfrac 12 (q_{11}^\dagger + q_{22}^\dagger),
\ -\tfrac i2\, (q_{12}^\dagger + q_{21}^\dagger) \right) \nonumber
\\
\vec \pi&=&(\vec \pi^\dagger)^\dagger
\quad D^\dagger = p^\dagger_{12}
\quad D = p_{12}
\quad M=\tfrac12 (n-\Omega) \nonumber
\end{eqnarray} 
where $Q_+$ generates charge density waves,
$\vec S$ is the spin operator, $\vec {\cal Q}$ is the staggered magnetization,
and $\vec \pi^\dagger,
 \vec \pi$ are the triplet D-wave pairs (Ref.\  \cite{zha97}), the operators
$D^\dagger, D$ are associated with singlet D-wave pairs, 
$n$ is the electron number operator,
and
$M$ is the charge operator. 
As we justify below, 
$SU(2)\tsub{p}$
is associated with
superconductivity, 
$SO(4)$ 
with AF order, 
and SO(5) 
with a ``spin-glass'' phase.

The group $SU(4)$  has a Casimir operator
\begin{equation} 
C_{su(4)}=\vec \pi^\dagger \hspace{-2pt}\cdot
\hspace{-2pt}\vec \pi + D^\dagger D +
\vec S \hspace{-2pt}\cdot\hspace{-2pt} \vec S + \vec {\cal Q}
\hspace{-2pt}\cdot\hspace{-2pt} \vec {\cal Q} + M(M-4)
\label{csu4}
\end{equation} 
and the irreps may be labeled by 3 quantum numbers,
$(\sigma_1,\sigma_2,\sigma_3)$. We choose a model
collective subspace
\begin{equation}
\ket{S} \equiv \ket{n_x n_y n_z n_d} = (\pi_x^\dagger)^{n_x}
(\pi_y^\dagger)^{n_y} (\pi_z^\dagger)^{n_z} (D^\dagger)^{n_d}
\ket{0}
\end{equation} 
which is associated with irreps of the form
$ (\sigma_1,\sigma_2,\sigma_3) = (\tfrac \Omega2,0,0).
$ The corresponding expectation value of the  $SU(4)$ Casimir
is a constant, 
$\ev{C_{su(4)}}=\tfrac\Omega2(\tfrac\Omega2 + 4)$. 

The full space is of dimension $2^{2\Omega}$ but
the collective subspace is much smaller, scaling as
$\sim \Omega^4$:
\begin{equation}
{\rm Dim\thin} (\tfrac{\Omega}{2},0,0) = \tfrac{1}{12}
(\tfrac\Omega 2 + 1)(\tfrac \Omega 2 + 2)^2
(\tfrac \Omega 2 + 3)
\label{eq12}
\end{equation}
This corresponds to truncation by a factor
$10^{54}$ for $\Omega \sim 100$.  For small lattices
we may enumerate all states of the collective subspace,
yielding a simple
model where observables can be calculated analytically.

The charge density wave operator
\begin{equation} 
Q_+= \sum_k (c_{k+Q\uparrow}^\dagger c_{k\uparrow} +
c_{k+Q\downarrow}^\dagger c_{k\downarrow}) =\sum_{rj}(-)^r n_{rj}
\end{equation} 
(where $n_{rj}$ is the number operator for electrons on lattice
site $r$ with spin
$j$) generates the $U(1)$ factor in $U(4) \rightarrow U(1) \times SU(4)$ and
commutes with all generators. Thus 
$\mel S{Q_+}S = 0$ and
charge-density waves are excluded from the present collective subspace. 

An $SU(4)$ model Hamiltonian can be constructed from a linear
combination of Casimir operators for all subgroups:
$H =A_0+\sum_{G_i}A_{G_i}C_{G_i}$,
where $A_0$ and $A_{G_i}$ are parameters and  the Casimir
operators $C_{G_i}$ are 
\begin{eqnarray}
C_{so(5)}\hspace{6pt}& =&\vec \pi^\dagger \cdot \vec \pi + \vec S \cdot \vec S,
\hspace{12pt} C_{so(4)} = \vec {\cal Q} \cdot \vec {\cal Q} + \vec S
\cdot\vec S \nonumber \\
C_{su(2)\tsub{p}} &=&D^\dagger D +M(M-1),\hspace{12pt}C_{su(2)\tsub s}
= \vec S \cdot \vec S
\nonumber \\
C_{U(1)}\hspace{7
pt}&=&M \mbox{ and }M^2
\label{casimirs} 
\end{eqnarray}
For conserved electron
number the terms in $M$ and
$M^2$ in Eq.\ (\ref{casimirs}) are constant.
Since $\ev{C_{su(4)}}$ is a constant, 
by using Eq.\ (\ref{csu4})  we can eliminate 
the $\vec \pi^\dagger \hspace{-2pt}\cdot
\hspace{-2pt}\vec \pi$ term and
by renormalizing the interaction strengths 
the $SU(4)$ Hamiltonian 
can be written
\begin{equation} 
H = H_0-G [\ (1-p)D^\dagger D + p\vec {\cal Q}\hspace{-2pt} \cdot
\hspace{-2pt}\vec {\cal Q}\ ] + \kappa\tsub{eff}\ \vec S
\hspace{-2pt}\cdot\hspace{-2pt} \vec S 
\label{generalH} 
\end{equation} 
with  $(1-p)G=G^{(0)}\tsub{eff}$, $pG=\chi_{\tsub{eff}}$ and
$\kappa\tsub{eff}$ as the effective interaction strengths, and $0\leq p \leq
1$. The term $H_0$ is a quadratic function of $n$ and may be parameterized as
$ 
H_0 =\epsilon\tsub{eff} \ n +\mbox{v}\tsub{eff} \ n(n-1)/2,
$
where $\epsilon\tsub{eff}$ and $\mbox{v}\tsub{eff}$ are
the effective  single-electron energy and the average
two-body interaction in zero-order approximation, respectively.   

In the full $SU(4)$ symmetry,
D-wave pairing,
antiferromagnetism, and $\pi$ collective modes
enter on an equal
footing.  
The system has
formed local $SU(4)$ pairs, which fixes the length of
vectors [$SU(4)$ Casimir] but not their
direction.  Physically, the system is
paired with $SU(4)$ symmetry but is neither
superconducting nor antiferromagnetic since the energy difference 
in those directions
are small on a scale set by the temperature.
Neither AF nor SC order parameters have finite
expectation values but a sum of their squares [the $SU(4)$
Casimir] does.

At this ``unification'' level, there is no distinction
among these degrees of freedom, just as in the Standard Electroweak Theory
the electromagnetic and weak interactions are unified above the intermediate
vector boson mass.  The full  symmetry should  hold while the
temperature of the system is sufficiently high that anisotropic
fluctuations in the directions associated with these collective degrees of
freedom are negligible, but not so high that thermal fluctuations destroy the
local $SU(4)$ pairs.
However, $SU(4)$ symmetry is broken to its subgroups at lower temperature,
leading to
3 dynamical symmetry chains with  eigenstates labeled by the length of
the $SU(4)$ vectors, but with orientations
no longer isotropic in the full
$SU(4)$ space.
These low-temperature
degrees of freedom may be interpreted as follows.

The three dynamical symmetry limits $SU(2), SO(4)$ and $SO(5)$, correspond
to the choices $p = 0$, 1, and 1/2, respectively,  in Eq.\ (\ref{generalH}).  
The Hamiltonian, eigenfunctions, energy spectrum and the corresponding quantum 
numbers of these symmetry limits are listed in Table I
(where we introduce a doping parameter $x$ that is 
related to particle number and lattice degeneracy 
through $x=1-n/\Omega$). 
The pairing gap $\Delta$ and the staggered magnetization
$Q$,
\begin{equation}
\Delta=\langle D^\dagger
D\rangle^{1/2}
\qquad
Q=\langle\vec{\cal Q} \hspace{-2pt}\cdot\hspace{-2pt}
\vec{\cal Q}\rangle^{1/2},
\end{equation}
may be used to characterize the states in these symmetry limits.
As we shall see, each 
limit represents a  
different possible low-energy phase of the 
$SU(4)$ system.

(1) The dynamical symmetry chain
$ SU(4) \supset SU(2)\tsub{p} \times SU(2)\tsub{s} \supset 
SU(2)\tsub{s} \times U(1)
$ corresponds to SC order and is the
$p=0$ symmetry limit of Eq.\ (\ref{generalH}).
The seniority quantum number $v$ is the number of
particles that do not form D-pairs. 
The ground state 
has
$v=0$, implying that all electrons are singlet-paired. In addition, there
exists a large pairing gap 
$\Delta = 
\tfrac12 \Omega (1-x^2)^{1/2}$,  
and $Q=0$. 
Thus we propose that this state is a
D-pair condensate, associated with the SC phase of the cuprates. 

(2) The chain
$ SU(4) \supset SO(4) \supset SU(2)\tsub{s}
$ corresponds to long-range AF order.  
This is the symmetry limit of Eq.\ (\ref{generalH})
when $p=1$.  
The $SO(4)$ subgroup is locally isomorphic to $SU(2)_F\times SU(2)_G$
generated by 
\begin{equation}
\vec{F}=\tfrac 12 (\vec{S}+\vec{\cal Q})
\qquad
 \vec{G}=\tfrac 12 (\vec{S}-\vec{\cal Q}), 
\end{equation}
where $\vec{F}$ and $\vec{G}$ are the total spin of
electrons at even sites and odd sites, respectively. 
Therefore, the $SO(4)$ Casimir
operator can be expressed as 
\begin{equation}
C_{so(4)} =2(\vec{F}\ ^2 +\vec{G}\ ^2).
\end{equation}
The
$SO(4)$ representations can be labeled 
by the spin-like quantum numbers 
$( F=w/2, G=w/2)$ and are of dimension $(w+1)^2$. The ground state
corresponds to $\omega=N$ and $S=0$, and has 
$n/2$ spin-up electrons on the
even sites ($F=N/2$) 
and $n/2$ electrons on odd sites with spin down
($G=N/2$), or vice versa. 
Thus it has maximal staggered magnetization 
\begin{equation}
Q=\tfrac12 \Omega(1-x)= \tfrac12 n
\end{equation}
and a large energy gap due to the 
correlation energy
$\vec{\cal Q}\hspace{-2pt}\cdot\hspace{-2pt}\vec{\cal Q}$ 
that inhibits electronic excitation, suggesting
magnetic insulator
properties at half filling. 
In addition, the pairing gap 
$\Delta =  
\tfrac12 \Omega(x(1-x))^{1/2}$ is small (zero at half
filling).
Thus we propose that these states are associated with an
AF phase of the cuprates.

(3) The dynamical symmetry chain
$ SU(4) \supset SO(5) \supset SU(2)\tsub{s} \times U(1) 
$ corresponds to a phase with spin-glass character.
This symmetry limit appears when $p=1/2$
in Eq.\ (\ref{generalH}) and has  
unusual
behavior. At half filling, $x=0$, the ground state is
highly degenerate with respect to the number of $\pi$-pairs $\lambda$, 
and 
mixing different numbers of $\pi$-pairs costs no
energy.  The $\pi$-pairs must be 
responsible for the antiferromagnetism in this phase, since 
only $\pi$-pairs carry spins. 
Thus the ground state 
in this symmetry limit has 
large-amplitude fluctuation in the AF order and  we 
propose that this symmetry limit 
may be associated with a spin-glass phase of the
cuprates.

Away from the symmetry limits there are no exact solutions
but the ground state properties can be studied easily using  
the coherent state method \cite{wmzha90}.
This is discussed in a separate paper \cite{lawu99}, but we
quote one result of that study here to reinforce our point concerning the
interpretation of the $SO(5)$ symmetry.
In Fig.\ 1, energy surfaces for the ground states 
for various particle number $n$ or doping $x$ are plotted   
as a function of a quantity $\beta$, which is related directly to the
AF order parameter (see figure caption). Three plots are associated
with the symmetry limits discussed above $(p=0, 1, 1/2)$ and one corresponds to
a situation with a slight $SO(5)$ symmetry breaking ($p= 0.52$). 

The energy minimum lies at 
$\beta = 0$ for all the $n$'s (or $x$'s) for the $SU(2)$ (Fig.\ 1a), and
at $\beta = [(1-x)/4]^{1/2}$ for the $SO(4)$ limit (Fig.\ 1b).
In Fig.\ 1c,
there is a broad range of doping in which the $SO(5)$ energy surface is
almost flat in the parameter $\beta$,
implying large excursions in the 
AF order.  Therefore,
the $SO(5)$ symmetry may be interpreted as a
phase having much of the character of a spin glass
for a range of particle numbers.
Notice in Fig.\ 1d that as doping varies the
$SO(5)$ Hamiltonian with slight symmetry breaking  
interpolates between AF order at half filling
and
SC order at smaller filling.
Thus $SO(5)$ acts as a kind of doorway between $SU(2)\tsub p$ and $SO(4)$
symmetries and
this gives a precise meaning to the
assertion \cite{zha97}
that the $SO(5)$ symmetry rotates the AF and
SC order parameters into each other.  

Our $SU(4)$ model and  Zhang's $SO(5)$ model \cite{zha97} 
have the same building blocks [the operator set (3)]. 
The essential
difference is that we implement the full quantum dynamics 
(commutator algebra) of these
operators exactly, while in Ref.\ \cite{zha97}, the dynamics is
implemented in an approximate manner: 
a subset of 10 of the operators 
acts as a rotation on the remaining 5 operators
$\{D^\dagger,D,{\cal \vec{Q}}\}$, which are treated phenomenologically
as 5 independent  
components of 
a vector.  

The embedding of $SO(5)$ in our larger algebra has various physical consequences
that do not appear if the $SO(5)$ subgroup is considered in isolation.
We list three:
(1)  A phase transition  from AF to SC 
at zero temperature and controlled by the
doping emerges naturally, 
whereas in the $SO(5)$ model a symmetry-breaking term proportional to
a chemical potential 
has to be introduced by hand.  
(2)  The
present results suggest that the $SO(5)$ subgroup 
is the appropriate description of
the underdoped regime, but that the AF phases at half filling and the optimally
doped superconductors 
are more simply described by our $SO(4)$ and $SU(2)\tsub p$
symmetries, respectively.  
(3)  As we shall discuss in a separate paper \cite{sun99},
the present $SU(4)$ theory leads naturally to the appearance of a pseudogap 
in the underdoped region \cite{TS99},
which occurs above the SC transition temperature $T_c$ 
and merges with $T_c$
near  the optimal doping point. 
 
The temperature dependence of the phase diagram requires
thermodynamic calculations that are 
in progress and will be presented separately.  
However, we may use the preceding discussion to
construct the  qualitative phase diagram
illustrated in Fig.\ 2.
First, 
$H_0$ in the Hamiltonian (\ref{generalH})
 may be regarded as the energy scale for the
$U(4)\rightarrow U(1)\times SU(4)$ symmetry,  representing 
a fermion system in which
electrons are all paired but with no distinction among the D-pairs and
$\pi$-pairs. 
We may
expect this symmetry to hold while the thermal energy is  
less than $H_0$. 
When the system is cooled, the
pairing and AF correlations [$H - H_0$, 
see Eq.\ (\ref{generalH})] become important,
$SU(4)$ breaks to its subgroups, 
and different low-temperature phases will appear
depending on the competition
between $D^\dagger D$ and $\vec
{\cal Q}\hspace{-2pt}\cdot\hspace{-2pt} \vec {\cal Q}$ 
interactions as a function of  
doping concentrations.  From the preceding discussion, at zero temperature we
expect the $SO(4)$
AF phase to dominate at half filling, the $SU(2)\tsub p$ SC phase to be favored
at larger doping, 
and the intermediate doping region to be described naturally by
the $SO(5)$ spin-glass phase that interpolates between SC and AF behavior with
doping.  Thus, $SU(4)$ symmetry implies the schematic phase diagram of Fig.\ 2,
independent of detailed calculations.

In summary, 
an $SU(4)$ model of High-$T_c$ SC has been  proposed 
that contains three phases:
A SC phase identified with the
$SU(2)\tsub{p}$
dynamical symmetry, an AF phase  identified with the
$SO(4)$ dynamical symmetry, and an $SO(5)$
phase extremely soft
against AF fluctuations over a substantial doping fraction
that we term (loosely) a
spin-glass phase.
Realistic systems may mix these sub-symmetries while 
retaining an approximate $SU(4)$ symmetry.  Zero-temperature
phase transitions 
are shown to be driven 
by the competition between the D-wave pairing and
the staggered magnetization $\vec{\cal
Q}\hspace{-2pt}\cdot\hspace{-2pt}\vec{\cal Q}$ interactions, as controlled 
microscopically by
the hole-doping concentration.  As we shall
discuss in detail separately, this model leads naturally to the appearance of 
pseudogaps in the underdoped regime.
Thus, we propose that high $T_c$ behavior of the cuprates results from  a
$U(4)$ symmetry realized dynamically, and because this symmetry is microscopic
its physical interpretation is accessible to calculation.  This 
provides a possible understanding of the cuprate phase diagram and substantial
insight concerning recent $SO(5)$ theories of D-wave superconductivity.

We thank Eugene Demler, Pengcheng Dai, and Ted Barnes for useful discussions.
L. A. Wu was supported in part by the National Natural Science Foundation of
China.  C.-L. Wu is supported by the  National Science Council of ROC.

\baselineskip = 14pt
\bibliographystyle{unsrt}

\begin{figure}
\caption{Coherent state energy surfaces for 
Eq.\ (9). The staggered magnetization
$Q$ is related to $\beta$ by 
$\langle Q \rangle = 2\Omega \beta_0 (n/2\Omega -\beta_0^2)^{1/2}$,
where $\beta_0$ is the value
of $\beta$ at the stable point, so
$\beta$ measures the AF order.
}
\label{fig1}
\end{figure}

\begin{figure}
\caption{Schematic phase diagram for $U(4)$ symmetry based on Fig.\ 1d. 
The $H_0$ in Eq.\ (9) is expressed in terms of hole doping $x$ with
$x=1-n/\Omega$; 
$\epsilon'_{\mbox{\protect\scriptsize{eff}}}=
\epsilon_{\mbox{\protect\scriptsize{eff}}}\Omega$, and
v$'_{\mbox{\protect\scriptsize{eff}}}
=$v$_{\mbox{\protect\scriptsize{eff}}}\Omega^2/2$. }
\label{fig2}
\end{figure}

\onecolumn
{\small
Table 1. The Hamiltonian, eigenstates and spectra in three dynamical
symmetry limits of the $SU(4)$ model.
$E_{g.s.}$ is the ground state energy, $\Delta E$ the excitation energy,
$N = n/2$ the pair number, $x = 1-n/\Omega$, and $\kappa_{so4}=
\kappa_{\mbox{\scriptsize{eff}}}+\chi_{\mbox{\scriptsize{eff}}}$.
}
{\scriptsize
$$
\begin{array}{lll}
\hline\\
\vspace{4pt}
 \mbox{SU(2) limit:  }|\psi(SU2)\rangle=\ket{N,v,S,m_s}
&\mbox{SO(4) limit:  }|\psi(SO4)\rangle=\ket{N,w,S,m_S}
&\mbox{SO(5) limit:  }|\psi(SO5)\rangle=\ket{\tau,N,S,m_S}\\
\vspace{4pt}
 \ev{C_{su(2)_p}}=\tfrac14(\Omega-v)(\Omega-v+2)
&\ev {C_{so(4)}}=w(w+2),\quad w=N-\mu
&\ev{C_{so(5}}=\tau(\tau+3),\quad \Omega/2-\tau=N-\lambda \\
 H =H_0+\kappa\tsub{eff}\ \vec S\hspace{-2pt}\cdot\hspace{-2pt}\vec S
&H =H_0+\kappa\tsub{so4}\ \vec S\hspace{-2pt}\cdot\hspace{-2pt}\vec S
&H =H_0+\kappa\tsub{eff}\ \vec S\hspace{-2pt}\cdot\hspace{-2pt}\vec S \\
\vspace{4pt}
 \hspace{18pt}-G^{(0)}\tsub{eff}\left [C_{su(2)_p}-M(M-1)\right]
&\hspace{18pt}-\chi_{\tsub{eff}}C_{so(4)}
&\hspace{18pt}-G^{(0)}\tsub{eff}\left [C_{su(4)}+M-C_{so(5)}\right ] \\
\vspace{4pt}
 E_{g.s.}=H_0-\frac{1}{4}G^{(0)}\tsub{eff}\Omega^2(1-x^2)
&E_{g.s.}=H_0-\frac{1}{4}\chi\tsub{eff}\Omega^2(1-x)^2
&E_{g.s.}=H_0-\frac{1}{4}\chi\tsub{eff}\Omega^2(1-x)^2\\
\vspace{4pt}
 \Delta E=\nu G^{(0)}\tsub{eff}\Omega+\kappa\tsub{eff}\ S(S+1),\quad \nu=v/2
&\Delta E=\mu \chi_{\tsub{eff}}(1-x)\Omega+\kappa\tsub{so4}\ S(S+1)
&\Delta E=\lambda\ x G^{(0)}\tsub{eff}\Omega+\kappa\tsub{eff}\ S(S+1)\\
\vspace{4pt}
 \nu = N, N-1, \ldots 0;\hspace{5pt} S=\nu, \nu-2, \ldots 0 {\rm \ or \ } 1
\hspace{10pt}
&\mu=N,N-2,\ldots 0 {\rm \ or\ } 1;\hspace{5pt} S=w, w-1, \ldots 0
\hspace{10pt}
&\lambda=N,N-1,\ldots 0 {\rm \ or\ } 1;\hspace{5pt} S=\lambda,
\lambda-2, \ldots 0 \\
\hline
\nonumber
\end{array}
$$
}

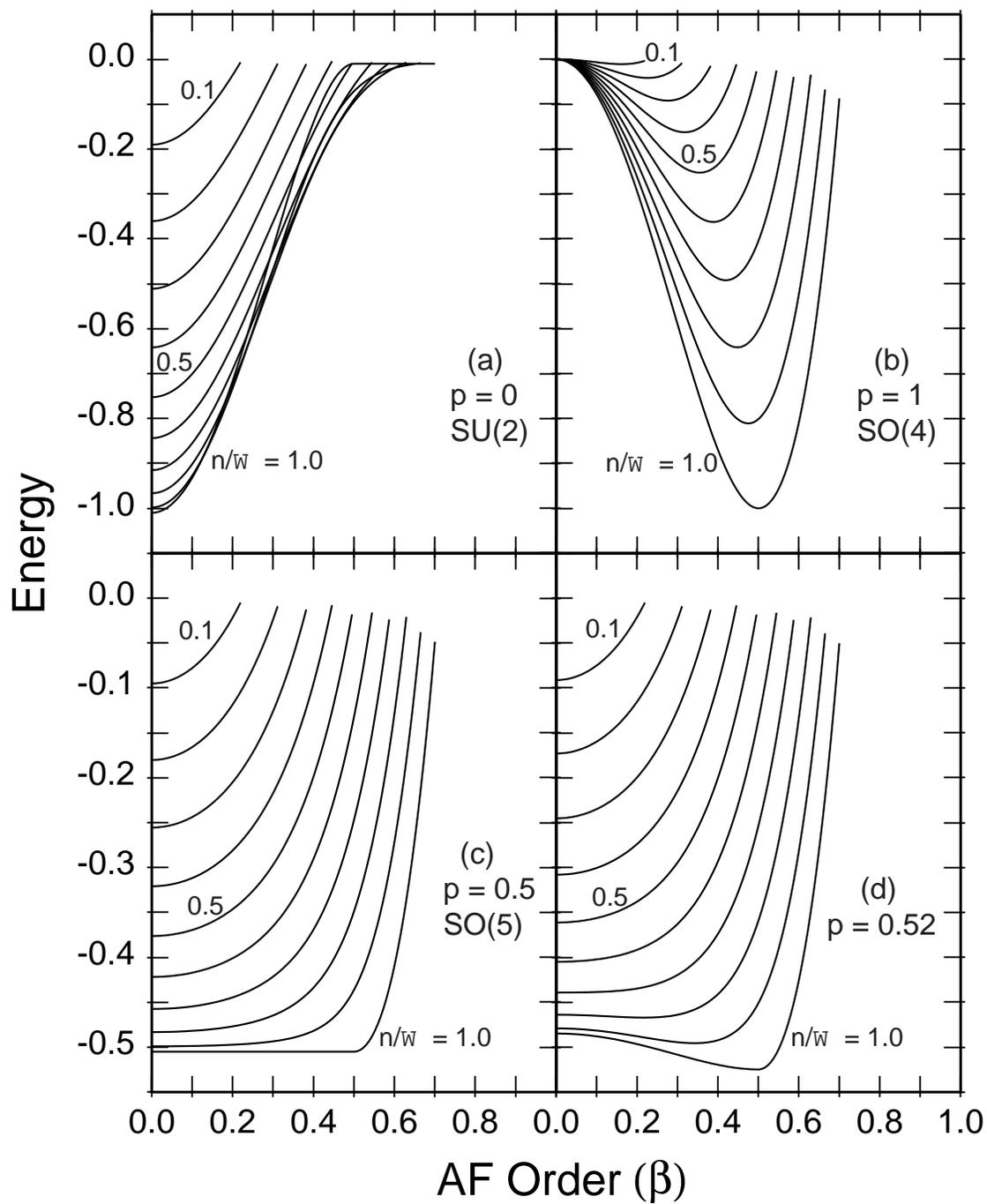
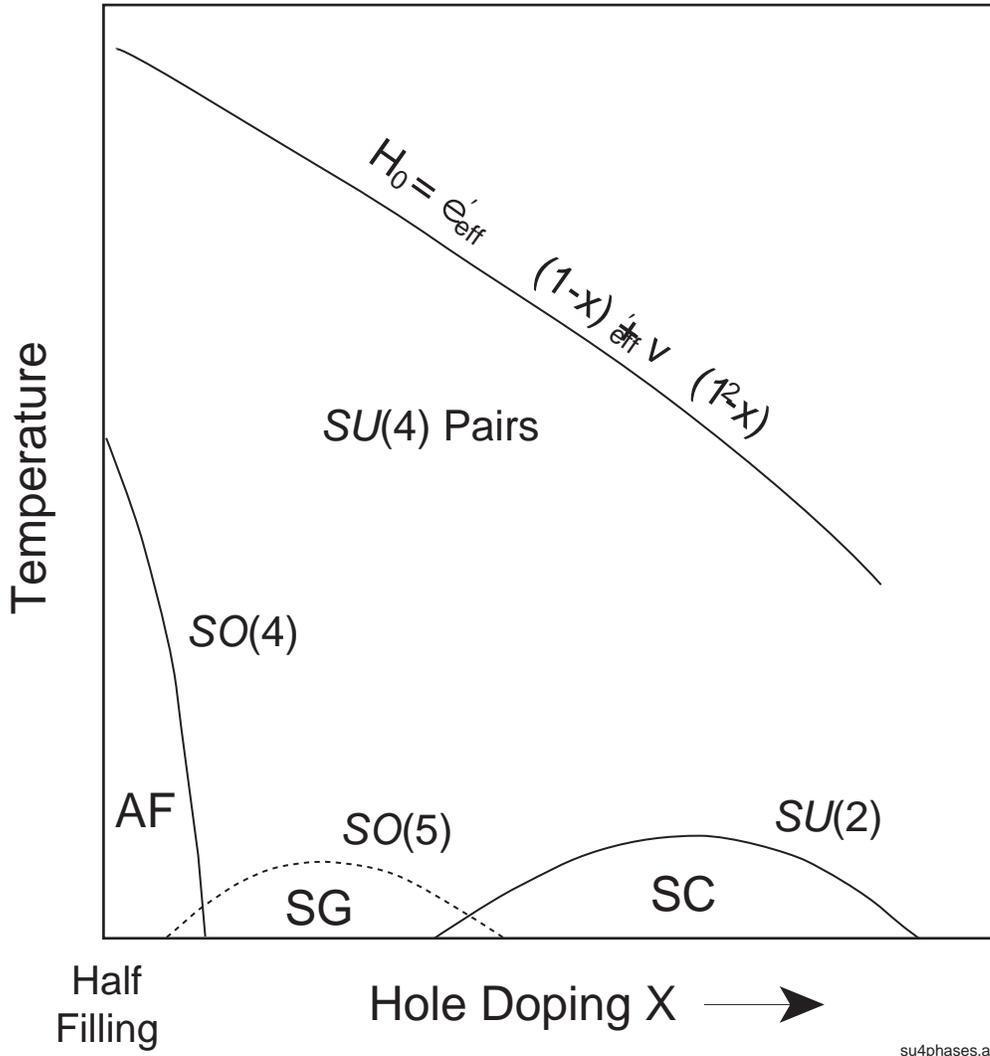
\end{document}